\documentclass[a4paper,11pt]{article}
\usepackage[]{amsmath,amssymb}
\usepackage{graphics,epsfig}
\newcommand{\be}{\begin{equation}}
\newcommand{\ee}{\end{equation}}
\newcommand{\beq}{\begin{equation}}
\newcommand{\eeq}{\end{equation}}
\newcommand{\bea}{\begin{eqnarray}}
\newcommand{\eea}{\end{eqnarray}}

\def\be{\begin{equation}}
\def\ee{\end{equation}}
\def\ba{\begin{eqnarray}}
\def\ea{\end{eqnarray}}

\begin{document}
\title {Entanglement entropy of a Maxwell field on the sphere}
\author{Horacio Casini and Marina Huerta \\
\it Centro At\'omico Bariloche and Instituto Balseiro\\
8400-S.C. de Bariloche, R\'\i o Negro, Argentina}
\maketitle
\vskip .5cm

\begin{abstract}
We compute the logarithmic coefficient of the entanglement entropy on a sphere for a Maxwell field in $d=4$ dimensions.
In spherical coordinates the problem decomposes into one dimensional ones along the radial coordinate for each angular momentum. 
 We show the entanglement entropy of a Maxwell 
field is equivalent to the one of two identical massless scalars from which the  mode of $l=0$
has been removed. This shows the relation $c^M_{\log}=2 (c^S_{\log}-c^{S_{l=0}}_{\log})$ between the logarithmic coefficient in the entropy for a Maxwell field $c^M_{\log}$, the one for a $d=4$ massless scalar $c_{\log}^S$, and the logarithmic coefficient $c^{S_{l=0}}_{\log}$ for a $d=2$ scalar with Dirichlet boundary condition at the origin. Using the accepted values for these coefficients $c_{\log}^S=-1/90$ and $c^{S_{l=0}}_{\log}=1/6$ we get $c^M_{\log}=-16/45$, which coincides with Dowker's calculation, but does not match the coefficient $-\frac{31}{45}$ in the trace anomaly for a Maxwell field. 
We have numerically evaluated these three numbers $c^M_{\log}$, $c^S_{\log}$ and $c^{S_{l=0}}_{\log}$, verifying the relation, as well as checked they coincide with the corresponding logarithmic term in mutual information of two concentric spheres.  
\end{abstract}

\section{Introduction}
In four dimensions the entanglement entropy (EE) on a sphere for a conformal field theory (CFT) admits an expansion of the form
\be
S(R)=c_2\frac{R^2}{\epsilon^2}+c_{\log}\log{\frac{R}{\epsilon}}+S_0\,,
\ee
where $\epsilon$ is a short distance cutoff, and $R$ is the sphere radius. 
The logarithmic coefficient $c_{\log}$ is expected to be independent of regularization.  Arguments based on conformal invariance of the theory imply that $c_{\log}$ is generally given by the coefficient multiplying 
the Euler density in the trace anomaly  
\cite{solodukin, casinihuertamyers}. The early prove of this result for $3+1$ dimensions done 
by Solodukhin \cite{solodukin} relied on conformal invariance and the connection between the EE with the holographic entropy. Later, a proof 
of the connection of the logarithmic coefficient and the trace anomaly was given for any even dimensions in \cite{casinihuertamyers} 
using conformal mappings to express the EE on a sphere as thermal entropy in de Sitter space 
with a fixed value of the product between the temperature and the curvature radius.  

For free scalar and fermion fields in $3+1$ dimensions this was confirmed numerically and 
analytically \cite{casinihuerta, dowker, scalar} by explicit calculations.
On the other hand, for a Maxwell field, the explicit thermodynamic calculation in de Sitter space by Dowker reveals a different result \cite{dowker}. 

This mismatch together with subtleties found to define correctly the partition of the Hilbert space as a tensor product for lattice gauge models \cite{donnelly}, 
inspired the introduction of the algebraic approach in \cite{gauge, maxwell} (see also \cite{lattice}), where the entropy is associated to local 
gauge invariant operator algebras rather than regions. 
There are ambiguities on the details of the choice of algebra at the boundary of the region, and these lead to ambiguities in the entropy.
 Several works computing the EE for a Maxwell field using directly methods of the continuum have also pointed out subtleties on boundary details of the calculations \cite{Huang,DonnellyWall,theisen,bunch}. Some authors suggested these boundary details can change the logarithmic coefficient \cite{Huang,DonnellyWall,theisen}, and with the appropriate choice the mismatch with the anomaly might be healed.    
 
However, as pointed out in \cite{gauge}, the algebra ambiguities in the continuum limit are of the same kind as the ones affecting the EE for other fields. In particular, the mutual information (MI) does not suffer any ambiguity in the continuum limit. Therefore, if we use mutual information to compute the logarithmic coefficient for a sphere there is no issue of boundary details in the calculation. Moreover, there is no known way to select a specific choice of algebra from the model itself, without introducing external elements that would make the calculation non universal. Most probably, in QFT the universal meaning for parts in EE of vacuum state is always contained in mutual information.   

In this work, we explicitly compute the EE for a Maxwell field (using an algebra without center, see \cite{gauge}) and the mutual information. We find the logarithmic coefficient coincides with the number calculated in \cite{dowker} and differs from the anomaly.

We first, exploiting the spherical symmetry of the problem, reduce the problem to a one dimensional one which depends only 
on the radial coordinate, in the same spirit as the method introduced by Srednicki in \cite{Srednicki} for scalar fields in spheres. We found 
the case of a Maxwell field is equivalent
to two copies of a massless 
scalar field,
where the angular momentum mode $l=0$ has been removed. This identification automatically tells us the logarithmic coefficient is
$c^M_{\log}=2(c^S_{\log}-c^{S_{l=0}}_{\log})$ where $c^S_{\log}=-\frac{1}{90}$ corresponds to the logarithmic coefficient for a massless scalar in $3+1$ dimensions and $c^{S_{l=0}}_{\log}=\frac{1}{6}$
to the $l=0$ mode of a scalar field. This gives $c^M_{\log}=-\frac{16}{45}$ for the Maxwell field.

We have successfully tested these results numerically, computing the EE in the 
lattice for a scalar, the scalar zero mode, and the Maxwell field.
We find the same is true for the logarithmic coefficients computed with mutual information.
As a crosscheck, we have computed also the area coefficients in the entanglement entropy and the mutual information
finding a perfect accord with the ones reported previously in the literature \cite{scalar, review}. 

The paper is organized as follows. In the second section, we discuss the planar problem of infinite parallel planes. This is useful as a warm up exercise, and
 because already in the planar geometry
there is an equivalence between the Maxwell and two massless scalar fields. In fact, the EE in planar geometry does not distinguish between the two theories, both have the same universal coefficient. In this sense, the sphere is different. In the third section, we show both theories differ in the zero angular momentum mode 
which is subtracted in the Maxwell theory. In the fourth section 
we check our results numerically.  Finally, we briefly discuss interpretations of the anomaly mismatch and speculate on possible resolution. 

\section{Mutual information for parallel planes}

\begin{figure}
\centering
\leavevmode
\epsfysize=6cm
\bigskip
\epsfbox{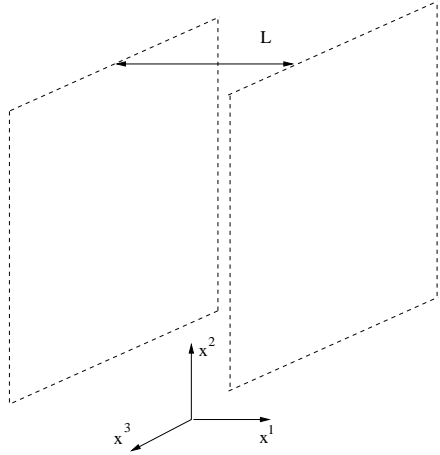}
\caption{Two parallel planes separated by a distance $L$ in the $x^1$ direction. These define the entangling surfaces for regions 
$A$ and $B$.}
\label{parallel}
\end{figure}

Before considering the problem of the EE for a Maxwell field on the sphere, we study the case two parallel planes separated by a distance $L$  as shown in figure (\ref{parallel}). Most of the ingredients in 
this discussion will be useful later for the spherical case.
The parallel planes define in turn
two regions $A$ and $B$ on each side
$A=\left\{x=\left(x^1,x^2,x^3\right): -\infty\leq x^1\leq 0\right\}$ and
$B=\left\{x=\left(x^1,x^2,x^3\right): L\leq x^1\leq \infty\right\}$.
We are computing the mutual information between these two regions. This is a finite and well defined quantity, given by the combination of entropies
\be
I(A,B)=S(A)+S(B)-S(A\cup B)\,.
\ee 
This case can be treated with dimensional reduction as discussed in \cite{review,scalarfield} for free scalar and fermions.

The Hamiltonian of the Maxwell field is  
\begin{equation}
H=\frac{1}{2}\int d^3x\, \left(E^2+B^2\right)\,,
\label{hammaxwell}
\end{equation}
with commutation relations
\be
[E_i(x),B_j(y)]=-i \epsilon_{ijk} \partial_k \delta(x-y)\,,
\ee
and constraints 
\be
\nabla E=\nabla B=0\,. 
\ee

We choose the planes perpendicular to $x^1$.  
In order to analyze the mutual information for this configuration we decompose the fields in Fourier sum in the two directions parallel to the plane. 
We assume the directions $x^2,x^3$ are compactified to large sizes $R_2,R_3$ with periodic boundary conditions. Writing $x\equiv x^1$ and $y=(x^2,x^3)$, we have 
\bea
E_i(x,y)=\sum_k \frac{e^{i k y}}{\prod_i(2\pi R_1R_2)^{\frac{1}{2}}} E_i(x,k)\,,\\
B_i(x,y)=\sum_k \frac{e^{i k y}}{\prod_i(2\pi R_1R_2)^{\frac{1}{2}}} B_i(x,k)\,.
\eea
Here the two component vector $k=(2 \pi n_2/R_2,2 \pi n_{3}/R_{3})$ where $n_2,n_{3}$ are integers, and the sum is over these integers. We also have
\bea
E_i(x,-k)=E_i(x,k)^\dagger\,,\\
B_i(x,-k)=B_i(x,k)^\dagger\,.
\eea 

Let us further decompose the vector components into the ones parallel and orthogonal to $k$,
\be
E_\parallel=\hat{k}.E\,,\hspace{0.3cm} E_\perp=(\hat{x}\times \hat{k}).E\,,\hspace{0.3cm}B_\parallel=\hat{k}.B\,,\hspace{0.3cm} B_\perp=(\hat{x}\times \hat{k}).B\,.\label{9}
\ee

The constraint equations tell that $E_\parallel$ and $B_\parallel$ are dependent operators  
\be
E_\parallel =\frac{i}{|k|} \partial_1 E^1\,,   \hspace{.5cm} B_\parallel =\frac{i}{|k|} \partial_1 B^1\,.\label{11}
\ee

The commutation relations decompose independently in each of the modes of fixed vector $k$. The non zero ones are
\bea
\left[ E_1(x,k),B_\perp^\dagger(x^{\prime},k^\prime)\right] &=& - |k| \delta(x-x^\prime) \delta_{k k^\prime}\,,\\
\left[ E_\perp(x,k),B_1^\dagger(x^{\prime},k^\prime)\right] &=& |k| \delta(x-x^\prime) \delta_{k k^\prime}\,.
\eea

The algebra of operators $E_i(x,k),E_i^\dagger(x,k),B_i(x,k),B_i^\dagger(x,k)$ is the same for $k$ and $-k$. We will write $\tilde{k}$ for the set $k,-k$ taken as equivalence class. 
Making the identifications
\bea
\phi_1=-\frac{i}{|k|} B_1\,,\hspace{.5cm} \pi_1=E_\perp\,,\\
\phi_2=-\frac{i}{|k|} E_1 \,,\hspace{.5cm} \pi_2=B_\perp\,, 
\eea
we have canonical commutation relations for the complex scalar fields $\phi_1$ and $\phi_2$ and their conjugate momentum. The Hamiltonian writes in these variables
\be
H=\sum_{\tilde{k}}\int d^2x\,\,\left(\pi_1^\dagger \pi_1+\pi_2^\dagger \pi_2+ k^2 \phi_1^\dagger \phi_1+k^2  \phi_2^\dagger \phi_2+\partial_1 \phi_1^\dagger \partial_1 \phi_1+\partial_1 \phi_2^\dagger \partial_1 \phi_2\right)\,.
\ee
This is precisely the dimensional reduction of two real scalar fields (see \cite{area}). Hence, as the local operator algebras and states are identical, 
mutual information for the Maxwell field in the wall geometry is given by twice the one for a massless four dimensional scalar field.
This last is in turn the sum over the mutual informations for the tower of massive one dimensional scalars. For a scalar field the final result is
\be
I=\kappa \frac{A}{L^2}\,,
\ee 
where $A$ is the wall area, $L$ the separating distance between the planes and $\kappa$ was computed in \cite{review, scalarfield}
\be
\kappa=  \left( (d-1)\,2^{d-2}\,\pi^{\frac{d-1}{2}}\,\Gamma\left(\frac{d-1}{2}\right)\right)^{-1} \int_0^\infty dy \,\,y^{d-2}\, c(y)=0.0055351600\hdots\,,
\ee
 with $c(r)$ the one dimensional entropic c-function.
The ``strip'' term for a Maxwell field is then twice the one for scalars. However, differences between the Maxwell field and 
massless scalars will show up for curved entangling surfaces.
  
\section{Entanglement entropy for a Maxwell field in the sphere}

We consider now the EE for a Maxwell field in the sphere. As before, the problem can be again dimensionally reduced, this time due 
to the spherical symmetry.

\subsection{Maxwell field: The Hamiltonian, constraints and commutators.}

In spherical coordinates, the vectors  $\bar{E}$ and $\bar{B}$ can be expressed as
\begin{equation}
 \bar{E}=E_{lm}^r(r)\bar{Y}^r_{lm}(\theta,\phi)+E_{lm}^e(r)\bar{Y}^e_{lm}(\theta,\phi)+E_{lm}^m(r)\bar{Y}^m_{lm}(\theta,\phi)\,,
\end{equation}
where the vector spherical harmonics $\bar{Y}^s_{lm}$ are defined in terms of the standard $Y_{lm}$ as 
\begin{eqnarray}
\bar{Y}_{lm}^r&=&Y_{lm}(\theta,\phi)\hat{r}\,,\,\,\,\,\,l\geq0\,,-l\leq m \leq l\,,\\
\bar{Y}_{lm}^e&=&\frac{r\bar{\nabla}Y_{lm}}{\sqrt{l(l+1)}}\,,\,\,\,\,\,l>0\,,-l\leq m \leq l,\\
\bar{Y}_{lm}^m&=&\frac{\bar{r} \times \bar{\nabla} Y_{lm}}{\sqrt{l(l+1)}}\,,\,\,\,\,\,l>0\,,-l\leq m \leq l,
\end{eqnarray}
and  satisfy the following orthogonality conditions
\begin{equation}
\int \bar{Y}^{s^{\prime}}_{l^{\prime}m^{\prime}}.\bar{Y}^{s \ast}_{lm} d\Omega=\delta_{s,s^{\prime}}\delta_{l,l^{\prime}}
\delta_{m,m^{\prime}}\,\,\,\,\,\,\,\,\,\,\,s,s^{\prime}=r,e,m\,\,.
\end{equation}
From there
\begin{equation}
 E^s_{lm}=\int \bar{E}.\bar{Y}^{s\,*}_{lm}d\Omega\,\,\,\,\,\,\,s=r,e,m\,\,.
\end{equation}
In this coordinates, the Hamiltonian (\ref{hammaxwell}) simply results
\begin{equation}
H=\sum_{lm} H_{lm}\,,\,\,\,\,\,\,\,\,l\geq0.
\end{equation}
with
\be
H_{lm}=\frac{1}{2}\int r^2 dr \sum_{s=r,e,m}\left[(E_{lm}^s(r))^2+ (B_{lm}^s(r))^2\right]\,,\,\,\,\,\,\,\,\,l>0\,,\label{hamiltonian}
\ee
and
\be
H_0=\frac{1}{2}\int r^2 dr \left[(E_{0}^r(r))^2+ (B_{0}^r(r))^2\right]\,.
\ee
The constraints tell the radial and electric component are not independent. From $\nabla.E=0$ and $\nabla.B=0$ we have
\begin{eqnarray}
 \frac{\partial E_{lm}^r}{\partial r }+\frac{2}{r}E_{lm}^r&=&\frac{\sqrt{l(l+1)}}{r}E_{lm}^e\,,\label{constraint1}\\
 \frac{\partial B_{lm}^r}{\partial r }+\frac{2}{r}B_{lm}^r&=&\frac{\sqrt{l(l+1)}}{r}B_{lm}^e\,,
 \label{constraint2}
\end{eqnarray}
where we have used that
\begin{eqnarray}
 \bar\nabla.\bar{Y}_{lm}^r&=&\frac{2}{r} Y_{lm}\,,\\
 \bar\nabla.\bar{Y}_{lm}^e&=&\frac{-\sqrt{l(l+1)}}{r} Y_{lm}\,,\\
 \bar\nabla.\bar{Y}_{lm}^m&=&0\,.
\end{eqnarray}
From here, it follows that $E_{lm}^e$ and $B_{lm}^e$ are dependent variables that can be written in terms of $E_{lm}^r$ and $B_{lm}^r$ respectively.
Eq.(\ref{constraint1}) fixes $E_{l=0}^r=\mathrm{const.}/r^2$. Since there are no charges, the only consistent solution finite at the origin 
 is $E_0^r=0$. 

Finally, we consider the commutation relations for the radial and magnetic components
\begin{equation}
 \left[E_{lm}^r(r),B^m_{l^{\prime} m^{\prime}} (r^{\prime})\right]= -\left[E^m_{lm}(r),B_{l^{\prime} m^{\prime}}^r(r^{\prime})\right]=\frac{\sqrt{l(l+1)}}{r^3}\delta(r-r^{\prime})
 \delta_{l,l^{\prime}}\delta_{m,m^{\prime}}\,.
\label{commutator1}
 \end{equation}
The other non zero commutators are 
the ones involving the dependent variables $E^e$ and $B^e$ which follow from the constraint equations. They will not be needed in what follows.
Replacing in the Hamiltonian (\ref{hamiltonian}) the constraint equations (\ref{constraint1}), (\ref{constraint2}) we obtain
\begin{equation}
 H_{lm}=\frac{1}{2}\int dr\, r^ 2 \left(\left(E_{lm}^r\right)^ 2+\left(B_{lm}^m\right)^ 2+\left(\frac{r}{\sqrt{l(l+1)}}\frac{\partial E_{lm}^r}{\partial r }
 +\frac{2}{\sqrt{l(l+1)}}E_{lm}^r\right)^2\right)+(E_{lm}\leftrightarrow B_{lm})\,.
\end{equation}
We can identify two identical sets of modes $(E^r, B^m)$ and  $(B^r, E^m)$.
Then, in order to reduce the commutation relations (\ref{commutator1}) to the canonical ones and to fix the coefficient of the square of the canonical 
conjugated momenta ($(E_{lm}^m)^2$ and $(B_{lm}^m)^2$) to $1$ in the Hamiltonian we introduce 
the following rescaled variables
\begin{eqnarray}
 \tilde{E}_{lm}^m&=&r E^m\,\,\,,\,\,\,\tilde{B}_{lm}^m=r B_{lm}^m\,,\\
 \tilde{E}_{lm}^r&=&\frac{r^2}{\sqrt{l(l+1)}}E_{lm}^r\,\,\,,\,\,\tilde{B}_{lm}^r=\frac{r^2}{\sqrt{l(l+1)}}B_{lm}^r\,.
\end{eqnarray}
The Hamiltonian and commutators in terms of the new variables read
\begin{equation}
  H_{lm}=\frac{1}{2}\int dr\left[ (\tilde{B}_{lm}^m)^2+\left(\frac{d\tilde{E}_{lm}^r}{dr}\right)^2+\frac{l(l+1)}{r^2}(\tilde{E}_{lm}^r)^2\right]
  + (\tilde{E}_{lm}\leftrightarrow \tilde{B}_{lm})\,,
 \label{hg2}
\end{equation}

\begin{equation}
\left[\tilde{E}_{lm}^r(r),\tilde{B}_{l^{\prime}m^{\prime}}^m(r^{\prime}) \right]=\delta(r-r^{\prime})\delta_{l,l^{\prime}}\delta_{m,m^{\prime}}\,.
\end{equation}
Note that $H_0=0$.

The boundary conditions in the origin for each mode $\tilde{E}_{lm}$ can be studied considering the classical equations. The Lagrangian, 
omitting the subscript $(lm)$, 
\begin{equation}
L=\frac{1}{2}\left(\dot{E}^2-{\tilde{E'}}^2-\tilde{E}^2 \frac{l(l+1)}{r^2}\right)\,,
\end{equation}
gives the following equation of motion
\be
\Ddot{\tilde{E}}(r,t)+\tilde{E}(r,t) \frac{l(l+1)}{r^2}-\tilde{E}''(r,t)=0\,.
\ee
For $\tilde{E}_{\lambda}(r,t)\sim e^{-i\lambda t}\tilde{E}_{\lambda}(r)$, we obtain
\be
\tilde{E}''_{\lambda}(r)-\tilde{E}_{\lambda}(r) \frac{l(l+1)}{r^2}+\lambda^2 \tilde{E}_{\lambda}(r)=0\,,
\ee
which gives $\tilde{E}_{\lambda}(r)=\sqrt{r}\left(C_1 J_{(l+\frac{1}{2})}(\lambda r)+C_2 Y_{(l+\frac{1}{2})}(\lambda r) \right)$.
Thus,  $\tilde{E}\sim r^{l+1}$ when $r\rightarrow 0$. We have set $C_2=0$ since this solution is divergent in the origin. 
If we think now in the original variables 
\be
E^r\sim \tilde{E}^r/r^2\sim r^{l-1}\,,\,\,\,\,\,\,\,\,\,B^m\sim \tilde{E}^r/r\sim r^{l}\,,
\ee
and
\be
E^e\sim r{E^r}'+2 E^r \sim r^{l-1}\,.
\ee
This tells us all the fields $\tilde{E},\tilde{B}$ goes to zero at the origin while the original ones can take a constant value for $l=1$.


\subsection{Scalar field}
The same analysis can be done for a scalar field. Using spherical coordinates, the radial Hamiltonian in three dimensions 
can be written as \cite{scalar,Srednicki}
\begin{equation}
H=\sum_{l=0}^{\infty}\sum_{m=-l}^{l} H_{lm}=\sum_{l=0}^{\infty}\sum_{m=-l}^{l}\frac{1}{2}\int_0^{\infty}dr\,\left(\tilde{\pi}_{lm}^2+
r^2\left[\frac{\partial}{\partial r}\left(\frac{\tilde{\phi}_{lm}}{r}\right)\right]^2+
 \frac{l(l+1)}{r^2}\tilde{\phi}_{lm}^2\right)\,,
\label{hscalar}
 \end{equation}
where $\tilde{\phi}_{lm}$ and $\tilde{\pi}_{lm}$ are defined in terms of the original field and momentum as
\begin{eqnarray}
 \tilde{\phi}_{lm}&=&r\,\int d\Omega \, \phi(r)Y_{lm}(\theta,\varphi)\,,\\
 \tilde{\pi}_{lm}&=&r\,\int d\Omega \, \pi(r)Y_{lm}(\theta,\varphi)\,.
\end{eqnarray}
such that 
\begin{equation}
 \left[\tilde{\pi}_{lm}(r),\tilde{\phi}_{l'm'}(r')\right]=i\delta(r-r')\delta_{ll'}\delta_{mm'}\,.
\end{equation}

Expanding the second term, we arrive at 
\begin{eqnarray}
 H_{lm}&=&\frac{1}{2}\int_0^{\infty}dr\,\left(\tilde{\pi_{lm}}^2+\left(\frac{\partial\tilde{\phi_{lm}}}{\partial r}\right)^2+\frac{l(l+1)}{r^2}\tilde{\phi_{lm}}^2
 -\frac{\partial}{\partial r}\left(\frac{\tilde{\phi_{lm}}^2}{r}\right)\right)\,, \nonumber\\
 &\equiv&\frac{1}{2}\int_0^{\infty}dr\,\left(\tilde{\pi_{lm}}^2+\left(\frac{\partial\tilde{\phi_{lm}}}{\partial r}\right)^2+\frac{l(l+1)}{r^2}\tilde{\phi_{lm}}^2\right)\,.\label{sham}
\end{eqnarray}
The boundary term $-\frac{\partial}{\partial r}\left(\frac{\tilde{\phi}^2}{r}\right)$ can be neglected since its corresponding boundary contribution vanishes  as 
$\sim r^{2l+1}$.

This Hamiltonian is identical to the Hamiltonian of each of the two electromagnetic spherical modes, eq. (\ref{hg2}), except for the additional $\phi_{l=0}$ radial mode. 
This $l=0$ mode is equivalent to a massless free scalar in $d=2$ 
with boundary condition $\phi_{l=0}(0)=0$ at the origin.
Thus, we conclude the problem for the Maxwell field in the sphere is equivalent to the one of two massless scalar fields where the $l=0$ mode has been removed.
In $2+1$ dimensions the identification between the algebra of the Maxwell 
field and the algebra of two truncated scalars follows directly from the duality $\frac{1}{2}\epsilon_{\rho\mu\nu}F^{\mu\nu}=\partial_{\rho}\phi$, 
and extends to any region \cite{maxwell}.

\subsection{Entanglement Entropy}

From the identification in the previous section, we conclude the Maxwell theory in the radial coordinate corresponds to 
two {\sl truncated} scalar fields with the $l=0$ mode removed. Due to the symmetry, the theory decouples in angular momenta such that
the total entropy  is written as an infinite sum of independent contributions $S=\sum_{l,m}S_{l,m}$. For a truncated scalar, the $l=0$ term 
is missing in the sum. 
We conclude that in particular the logarithmic coefficient in the EE
for a sphere must be 
\begin{equation}
 c^{M}_{\log}=2\left(c^{S}_{\log}-c^{S_{l=0}}_{\log}\right)\,,
 \label{logcoef}
\end{equation}
where $c^{S}_{\log}$ is the log coefficient for a $3+1$ dimensional scalar field in a sphere,  and $c^{S_{l=0}}_{\log}$ is the one of a one dimensional 
massless scalar in an interval $\left(0,R\right)$ with condition $\phi(0)=0$ at the origin.  
 Both, $c^{S}_{\log}=-1/90$ \cite{solodukin,casinihuerta} and $c^{S_{l=0}}_{\log}=1/6$ \cite{cc}, are known to correspond to the conformal anomalies of the associated theories.
 We have
\begin{equation}
 c^{M}_{\log}=2\left(-\frac{1}{90}-\frac{1}{6}\right)=-\frac{16}{45}\,.
\end{equation}
This is the value found by Dowker in \cite{dowker} by thermodynamical arguments in de Sitter space.

\section{The lattice realization for spherical sets}
We check numerically the results found above, evaluating the EE for Maxwell and  scalar fields in the sphere and the scalar zero angular momentum mode field in the line.
We start reviewing very briefly the techniques we are going to use (see \cite{review} for a review). 
Finally, we also consider the mutual information. All the numerical results confirm the ones discussed in the previous Sections.

\subsection{Entropy for scalar and gauge fields}

In general, for a set of fields $\phi_i$ and $\pi_i$ with canonical commutation relations, the entanglement entropy associated to a region $V$, 
can be calculated from the field and momentum
correlators $X=\left\langle \phi_i \phi_j\right\rangle$  and $P=\left\langle \pi_i \pi_j\right\rangle$ restricted to $V$ \cite{review}.  
These, in turn, are functions of the matrix $K$ 
\begin{equation}
 X_{ij}=\frac{1}{2}K^{-\frac{1}{2}}_{ij}\,,\,\,\,\,\,\,\,P_{ij}=\frac{1}{2}K^{\frac{1}{2}}_{ij}\,,
 \label{corr}
\end{equation}
defined from the discrete Hamiltonian
\begin{equation}
 H=\frac{1}{2}\left(\sum_{i} \pi_i^2+\sum_{ij}\phi_iK_{ij}\phi_j\right)\,.
 \label{matrixk}
\end{equation}

The entropy is written in terms of $C=\sqrt{\left.X\right|_V.\left.P\right|_V}$ as
\begin{equation}
 S=(C+1/2)\log(C+1/2)-(C-1/2)\log(C-1/2)\,.
\label{nentropy}
 \end{equation}

For spherical sets, the problem can be reduced to a one dimensional one in the radial coordinate as shown in the previous section. 
In our case, for each $l$ the Hamiltonian is

\begin{equation}
H_{l}=\frac{1}{2}\sum_{i}\left[\pi_i^2+\phi_i^2\frac{l(l+1)}{i^2}+\left(\phi_{i+1}-\phi_{i}\right)^2\right]\,,
\label{discretehamiltonian}
\end{equation}
which is simply the discrete version of the radial Hamiltonian (\ref{hg2}) in the previous section. More precisely, there are two 
identical and independent set of modes with this same Hamiltonian.
From (\ref{discretehamiltonian}), we identify the matrix $K^{l}$
\begin{eqnarray}
 K^{l}_{1,1}&=&l(l+1)+1\,,\\
 K^{l}_{i,i}&=&\frac{l(l+1)}{i^2}+2\,,\\
 K^{l}_{i,i+1}&=&K_{i+i,i}=-1\,.
\end{eqnarray}

We note, that this matrix is different from the one used by Srednicki \cite{Srednicki}. This is simply due to the fact that we are implementing a different discretization. 
This will not spoil the final continuum limit. In fact, both $K's$ give rise to the same correlator in the continuum. 
As a crosscheck, in the large lattice size limit, we have tested the correlators (\ref{corr})
tend to the ones in the continuum where $K^l$ can be directly read from (\ref{sham}) and corresponds 
to the operator $-\partial^2_r+\frac{l(l+1)}{r^2}$.
More explicitly, the eigenfunctions of $K^l$ satisfy
\begin{equation}
 \left(-\partial^2_r+\frac{l(l+1)}{r^2}\right)\psi_k(r)=k^2 \psi(r)\,,
\end{equation}
with two solutions
\begin{equation}
\psi_1(r)=\sqrt{r}J_{l+1/2}(kr)\,,\,\,\,\,\psi_2(r)=\sqrt{r}Y_{l+1/2}(kr)\,.
\end{equation}
We only keep the first one since the second one diverges in $r=0$. Then
\begin{equation}
 \psi_k(r)=\sqrt{kr}J_{l+1/2}(kr)\,,
\end{equation}
with a normalization prefactor such that $\int_0^{\infty}dr \psi_k(r)\psi_{k^{\prime}}(r)=\delta(k-k^{\prime})$.
For $r>r'$, this gives
\begin{equation}
\left\langle \phi_l(r) \phi_{l}(r')\right\rangle=\frac{1}{2}\int_0^{\infty}dk\,\psi_k(r)\frac{1}{k}\psi^*_k(r')=\left(\frac{r^{\prime}}{r}\right)^{l+1}\frac{\Gamma(l+1)}{\sqrt{\pi}}\, {_2}F_1 \left(1/2, 1 + l, 3/2 + l, r'^2/r^2 \right)\,.
\end{equation}
The lattice correlators approach this result for large $r,r'$.

The total entropy will be given by the sum of the contributions $S_l$ for each mode
\begin{equation}
 S=2\sum_{l=1}^{\infty}(2 l+1)S_{l}\,,
 \label{entropy}
\end{equation}
where $S_l$  depends on $C$ as in (\ref{nentropy}) and the factor of two counts for the two sets of modes $(B^m, E^r)$ and $(E^m, B^r)$.

\begin{figure}
\centering
\leavevmode
\epsfysize=6cm
\bigskip
\epsfbox{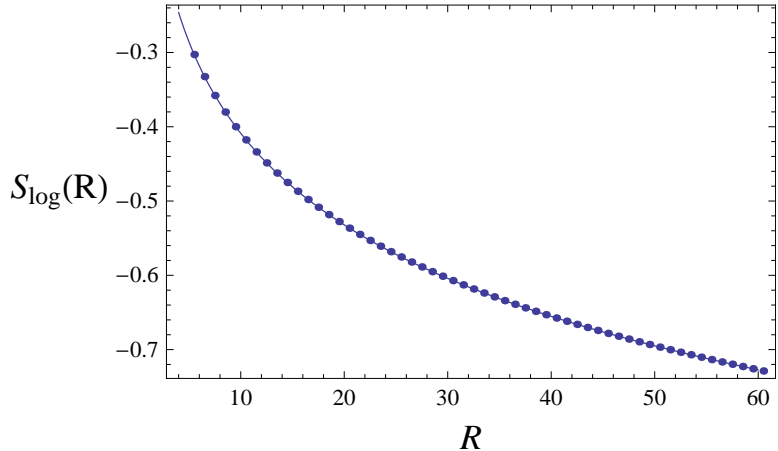}
\caption{The sphere entanglement entropy for a Maxwell field where we have subtracted the area and constant terms. The fitting curve is $0.17763 \log(R)$}
\label{entropymaxwell}
\end{figure}

The details of our numerical calculation are as follows. The sum of entropy contributions in (\ref{entropy}) has been calculated for a range of 
radius $n=5,...,60$ (measured in lattice sites) exactly up to $l_{max}=1000$.
The large $l>l_{max}$ contribution is calculated for each $R$ by fitting eight different values of $S_l$ 
from $l=1000$ to $l=4500$ \cite{scalar}. The total size of the radial lattice is given by a finite infrared cutoff $N$. We impose $\phi_l(N)=0$.
To eliminate the dependence on the infrared cutoff $N$, after summing over $l$, we repeat the calculations 
for different lattice sizes $N=200,300,400,500$ and obtain the infinite lattice limit fitting the results with $a_0+\frac{a_{-2}}{z^2}$ for each radius. 
We take $a_0$ as the infinite lattice limit.

Finally, we fit the entropy with $c_0+c_2 R^2+c_{\log}\log(R)$ where we define the sphere radius as $R=n+\frac{1}{2}$ in terms of the number of lattice sites. 
We obtain 
\begin{equation}
c^M_{\log}=2\times-0.17763 \sim -\left(\frac{16}{45}\right)=2 \times -0.1777\,.
\end{equation}
The results are shown in figure (\ref{entropymaxwell}).

As a crosscheck, we also measure the (non universal) area term. As expected, we obtain 
\begin{equation}
c_2=0.295431\,,
\end{equation}
which agrees with the same coefficient 
found for a scalar field in \cite{scalar} using a the discretization of Srednicki, up to six digits. 

We have also done the computation of the logarithmic coefficient for a massless scalar just removing the prefactor of two and adding 
the $l=0$ mode in (\ref{entropy}). We find $c^{S}_{\log}=-0.01116\sim -\frac{1}{90}=-0.1111\hdots$ consistent with the calculations in \cite{scalar}.

\begin{figure}
\centering
\leavevmode
\epsfysize=6cm
\bigskip
\epsfbox{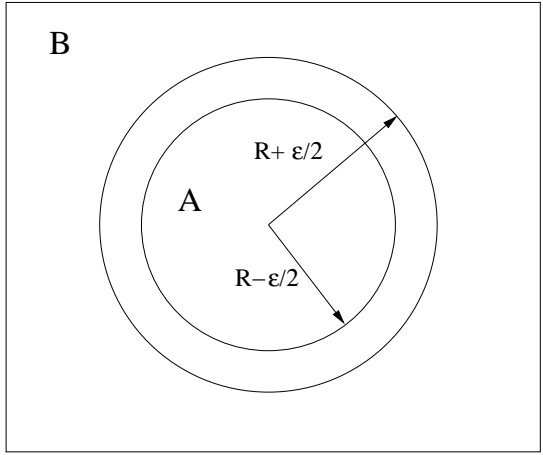}
\caption{Two sets, $A$ and $B$: $A$ is a sphere of radius $R_1=R-\epsilon/2$ and $B$, the complimentary region of a sphere of radius $R_2=R+\epsilon/2$. 
The averaged radius is $R=(R_1+R_2)/2$ and the annulus section is $\epsilon=R_2-R_1$.}
\label{set}
\end{figure}

\subsection{Mutual information }

Mutual information gives us a geometrical prescription for defining a universal 
regularized entanglement entropy \cite{fmutual}. Consider the geometry shown in figure (\ref{set}), the mutual information $I(A,B)$ we interested in is the one between
a sphere of radius $R_1$ and the complementary region of the sphere of radius $R_2$. The mutual information depends on the averaged radius $R=\frac{1}{2}(R_1+R_2)$ and the 
separation $\epsilon=R_2-R_1$. In the limit $\epsilon\rightarrow 0$, the {\sl regularized} entropy is defined as
\begin{equation}
S(R)=\frac{1}{2}I(R,\epsilon)\,.
\end{equation}
\begin{figure} [tbp]
\centering
\leavevmode
\epsfxsize=10cm
\bigskip
\epsfbox{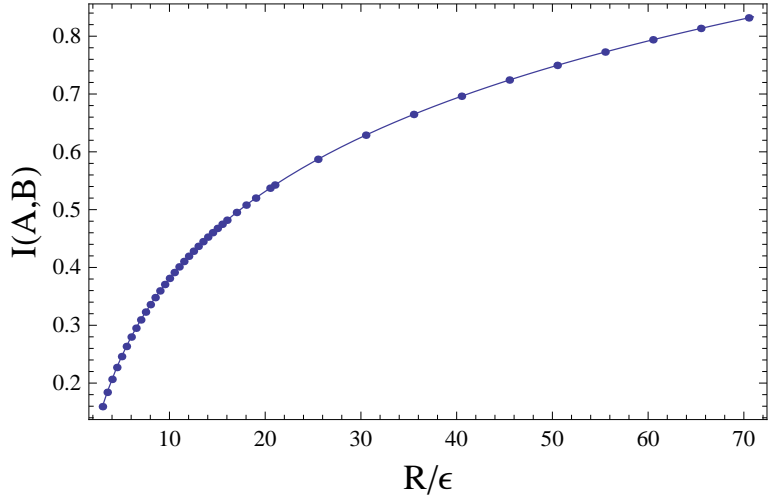}
\caption{Mutual information between $A$ and $B$  (Fig.(\ref{set})) for the scalar $l=0$ mode. The solid interpolating curve is $f(\eta)$. }
\label{zeromode}
\end{figure}
We use $S(V)=S(-V)$ for pure states to calculate $S(B)$ and $S(A\cup B)$, as the entropies associated to the sphere of radius $R_2$ and the annulus 
with inner and outer radius $(R_1,R_2)$ respectively. 

The mutual information calculation is more subtle numerically than the one for the entropy, since the log coefficient in the subtraction of disks and annular strip entropies 
is very sensitive to numerical errors. On the other hand, MI has the advantage to be less sensitive to ultraviolet contributions, what allows us to cut
the sum over angular momenta to smaller values.
By inspection, we have found the contribution from angular momenta vanishes as $l$ increases, being negligible for $l\geq 100$ already for the range of radius $R_1$ and 
$R_2$ we are using. 
For a lattice size $N=2000$ and $l_{max}=150$, we calculate the mutual information for different configurations with
\be
\eta=\frac{R}{\epsilon}=\frac{(R_1+R_2)}{2(R_2-R_1)}\,.
\ee
We take $\eta$ in the range $8 \hdots 22$.
The mutual information at each fixed value $\eta$ is evaluated for different $R$ and fitted with $a_0+a_2/R^2+a_4/R^4$ where $R=\frac{1}{2}(R_1+R_2)$.
The continuum limit for each $\eta$ is $a_0$.
\begin{figure} [tbp]
\centering
\leavevmode
\epsfxsize=10cm
\bigskip
\epsfbox{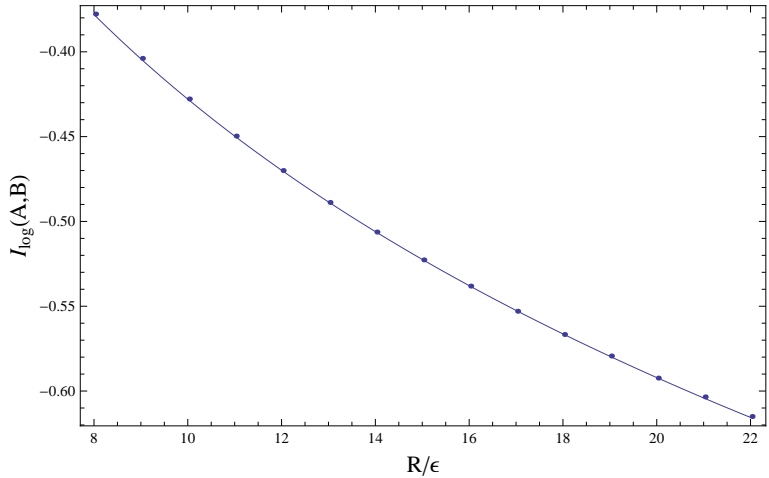}
\caption{Mutual information between $A$ and $B$ (Fig.(\ref{set})) for a Maxwell field for a single set of modes $(B^m, E^r)$. 
In the plot, the area term has been subtracted from the data. The fit shown is $-0.9899 (f(\eta)+\frac{2}{90}\log(\eta))$ with 
$f$ the interpolating function of the MI of the $l=0$ massless scalar (Fig.(\ref{zeromode}))}.
\label{mutual}
\end{figure}
In the continuum we expect, for large $\eta$,
\be
\frac{1}{2}I(\eta)=s_2 \eta^2 + s_{log} \log (\eta)+\text{subleading terms}\,.
\label{mutual12}
\ee
In order to gain precision in the computation of the logarithmic coefficient, we will profit of the knowledge of the theoretical value of the 
coefficient $s_2$ of the area term in the MI 
proportional to $\eta^2$. This coincides with the area term of the MI between two parallel planar entangling surfaces. 
This is calculated independently using an analytical dimensional reduction approach (see Section 2) and found to 
be $s_2= 4\pi\times\kappa$ with $\kappa = 0.0055351600$ \cite{review}. 
In fact, we can check numerically that fitting the data with a curve $s_2 \eta^2 + s_{log} \log (\eta)+s_0$ we obtain the numerical value $s_2=0.0695355$ 
consistent with the theoretical one. 

The fit to the numerical data is excellent for the area coefficient, but it is unstable and contains significant deviations from the expected result for the logarithmic ones.
For relatively small values of $\eta$ we are considering, it seems there are important subleading contributions that make the logarithmic coefficient unstable 
within $30\%$ error. 
In order to have a more stable fit, our strategy here is to profit from the fact that the main contribution to the logarithmic term for the Maxwell field as compared 
to the full scalar, comes from the 
removed $l=0$ scalar mode ($1/6$ compared to $1/90$). As it appears, this is also the main source of the subleading corrections.
Thus, we first subtract the area term $4\pi\kappa \eta^2$ to the data  and fit the result with 
\be
\frac{1}{2}I(\eta)-s_2 \eta^2= x \,(f(\eta)+\frac{2}{90}\log(\eta))+\text{const.}\,,
\ee
where $f(\eta)$ is an interpolating function corresponding to the 
MI for the zero angular momentum scalar mode (figure (\ref{zeromode})) and the $1/90$ corresponds to the contribution of the full scalar. This gives in fact a more stable fit 
for the Maxwell field. We get 
$x= -0.9899$. 

The MI for the $l=0$ mode has a logarithmic dependence in $\eta$ for large values of $\eta$, which is in accordance with the interpretation of
MI as a regularized entropy in this limit. However, it also contains a subleading $-1/2\log(\log(\eta))$ correction. 
This comes from a term $1/2\log(\log(\eta))$ in the entropy of the small interval of size $\epsilon$ as we put $R$ to infinity. This correction is related to the infrared divergences for a massless scalar in $d=2$, that here are regulated by the distance $R$ to the boundary. A corresponding divergence appears for a massive scalar in the limit of small mass \cite{review}. 
That is, we get 
\be
f(\eta)=\frac{1}{3}\log(\eta)-\frac{1}{2}\log(\log(\eta))+\textrm{const}\,,\hspace{1cm} \eta\gg 1\,.
\ee
 However, we note the approach to this regime 
is quite slow and this is the reason why a direct fit with (\ref{mutual12}) is not adequate.
Then, we can read off from the result of the fit for $x$ that for large $\eta$ we have
\begin{equation}
s_{log}=2(-0.9899)\left(\frac{1}{6}+\frac{1}{90}\right)=-0.3519 \sim -16/45=-0.35555\hdots\,.
\end{equation}
Therefore, as expected, we found that the logarithmic coefficient given by MI coincides with the one obtained directly from the entropy.
The results are shown in figure (\ref{mutual}). Notice that these results imply the mutual information for the Maxwell field contains also a subleading term $\log(\log(\eta))$ for large $\eta$. The complete expression reads
\be
\frac{1}{2}I(\eta)=s_2 \eta^2 -\frac{16}{45} \log (\eta)+\frac{1}{2}\log(\log(\eta)) +s_0\,.
\ee

\subsection{Algebras with center}
In gauge theories the most natural choices of local gauge invariant operator algebras contain a center, that is, a set 
of  operators commuting with all other operators in the algebra \cite{gauge}. This is generally the case when the number of electric and magnetic operators is not balanced. For example, 
the electric center \cite{gauge} corresponds to the case where there are a number of extra electric operators localized at the boundary. The entropy in the case of local algebras with center has an additional classical contribution. 
The calculation of $c^M_{\log}$ done in previous sections corresponds to the case of operator algebras without center, as follows 
from the matching between electric and magnetic degrees of freedom.
For the Maxwell field in $d=4$ the electric and magnetic centers choices are dual to each other and the calculations for these two choices are equivalent in the spherical lattice. For example, the electric center can be implemented by adding the operators $E_{lm}^r[n+1]$ on the boundary to the full algebra
of operators up to radius $n$. 
The classical entropy for this center can be computed using the formula for Gaussian states in \cite{gauge}. We have 
\be
S_{\text{clas}}=\sum_{l=1}^{\infty}(2l+1)\left(\log\left(\langle E_{lm}^r[n+1]^2\rangle\right)+\text{const.}\right)\,.
\label{clasica}
\ee
The constant is arbitrary because the definition of the classical entropy for continuous variables has an additive ambiguity.

We find numerically that 
\be
\log\left(\langle E_{lm}^r[n+1]^2\rangle\right)\sim -\log(l)
\ee
for $l\gg n$. Then, even disregarding the problem of the ambiguity in the definition of the classical entropy, we see that the sum in (\ref{clasica})
does not converge. This means that the radial discretization is not enough to regularize the entropy in this case.\footnote{Radial discretization is also known to be not enough to regularize the scalar entropy for dimension $d>4$ \cite{latorre}. We expect however that MI can be computed in a spherical lattice also for $d>4$.} 

As shown in \cite{gauge}, mutual information in the continuum limit is independent on the details of algebra choice. With or without center, it must converge to a unique universal value. In fact, the contribution of the classical center on the boundary to mutual information seems to vanish in the continuum limit\footnote{By monotonicity of MI, this must be the case if the mutual information of regions with width $\epsilon$ tending to zero vanishes in the $\epsilon \rightarrow 0$ limit. This is related to the question if there exist well defined operators in the theory which are smeared in a $d-2$ dimensional region (in contrast to the usual smearing with test functions having support in $d$ dimensional regions). }. In this sense, the calculation of the previous section 
has a universal character. Nevertheless, we have checked numerically that the classical contribution to the mutual information for two radius $R_1$ and $R_2$  vanishes exponentially fast as a function of $l$ for large $l>R_1,R_2$. This produces a finite contribution when summed over angular modes. That is, even if the entropy in the electric center case cannot be computed using radial discretization, there is no obstacle to compute mutual information. We also checked the classical contribution to MI decreases towards the continuum limit, when $R_1,R_2$ are taken large with respect to the lattice spacing. 

\section{Discussion}

Our main result is that for a free Maxwell field the logarithmic coefficient $c_M$ in the entanglement entropy of a sphere in $3+1$ dimensions 
is not given by the coefficient $-31/45$
multiplying the Euler density in the trace anomaly. It rather coincides with Dowker's result \cite{dowker}, and is given by 
\begin{equation}
 c^{M}_{\log}=2(c^S_{\log}-c^{S_{l=0}}_{\log})=-\frac{16}{45}\,,\label{llcc}
\end{equation}
with $c^S_{\log}=-1/90$ and $c^{S_{l=0}}_{\log}=1/6$ the logarithmic coefficients for a $d=4$ massless scalar and $d=2$ massless scalar with Dirichlet boundary condition at the origin. Our results in terms of the mutual information show this is a solid equivalence, that cannot be modified by local boundary changes in the algebra prescriptions. 

On the other hand, the logarithmic coefficient for the entropy proper, without invoking MI, can probably be tuned using particular algebras with center, and hence should not be universal. The standard understanding in this regard is that the full contribution for the Maxwell field 
has two parts, a bulk and a boundary contribution. This last one has been associated for example to an electric center \cite{DonnellyWall}, or equivalently to boundary degrees of freedom localized on the 
entangling surface generated by the gauge redundancy \cite{Huang}. With this choice, the contribution to the logarithmic coefficient would be the one of a {\sl ghost} massless scalar in $S^2$, 
\begin{equation}
c^{M}_{\log}=c_{bulk}+c_{boundary}=-\frac{16}{45}-c_{s}(S^2)=-\frac{16}{45}-\frac{1}{3}=-\frac{31}{45}\,.
\end{equation}
In the same spirit, in \cite{Astaneh} and \cite{theisen} the $-16/45$ is corrected to match the expected value correcting 
the effective action by relevant boundary terms or total derivatives.

In any case, for any such choices, mutual information would return the same logarithmic coefficient (\ref{llcc}) because MI by 
its very definition is insensitive to regularization dependent boundary terms. Therefore, we think the result (\ref{llcc}) is definitive. 
Most probably, there is no universal meaning for the entropy of QFT in Minkowski vacuum other than the one given by mutual information.

This leaves two open problems to which we hope to return in the future. The first one is why mapping the EE to the problem of the logarithmic 
coefficient of the free energy on a $d$-dimensional Euclidean sphere in \cite{casinihuertamyers} does not produce the right coefficient in this particular 
case. In this sense, we note this calculation actually computes a ``naked'' entropy, and a careful examination is necessary to determine how to modify it 
to obtain the result for the mutual information. This might be specially important for free bosonic models in which exist operators with  dimensions $d-2$ that 
can be added as surface terms to the modular Hamiltonian $K$. Boundary terms in the modular Hamiltonian will produce insertions on the Euclidean sphere 
equator. These boundary terms can be fixed by the first law for the variation of the entropy under infinitesimal variations of the 
state, $\delta \langle K\rangle=\delta S$, where $\delta S$ is understood as half the variation on mutual information for vanishing cutoff. A 
boundary term is known to appear in $K$ for the scalar field \cite{herait}. Any understanding in this line should account to the fact that the free energy 
on the sphere does give the right result for scalars. 

The second question is whether there is a sense in which the anomaly coefficient can be recovered as the correct result. We want to speculate that this 
might indeed be possible for charged theories, perhaps giving a universal meaning to the calculations in \cite{Huang,DonnellyWall}. As we have seen this 
is not possible for free Maxwell field. However, the situation might be different for a Maxwell field coupled to charges. If the charged fields are heavy 
with a large mass scale $M$, and we evaluate mutual information of concentric spheres with $RM\gg 1$, two different situations might appear depending on 
the value of $M\epsilon$. If $M\epsilon \gg 1$, then we expect massive modes cannot alter the result (\ref{llcc}). In this case the only connection between 
the two regions is through the massless Maxwell field with the usual free correlators. However, if $M\epsilon \ll 1$ and the scale of $\epsilon$ has crossed 
the scale of the masses, charged particle fluctuations will be visible to mutual information. In general massive particle fluctuations will contribute locally 
on the entangling surface to the area term, but in principle could not change the infrared $\log(R)$ term. However, here the charges will allow the 
constraint  $\nabla E=\rho$ to talk between the two regions.

 Can this produce a logarithmic coefficient given by the anomaly? Note that in this case the extra contribution has to be highly universal, 
 independent of the particle charges for example. This might seem odd, but a very similar situation is expected to hold for topological theories in $2+1$ dimensions \cite{fmutual}. In this case, the mutual information must be zero for $\epsilon M\gg 1$, where $M$ is the gap scale. This is because there are no correlations at distances larger than $\epsilon$. But as we put $\epsilon M\ll 1 $ correlations of the underlying physics should built a mutual information different from zero, and in particular, the constant term should give the topological entanglement entropy $-\gamma$ characterizing the topological order.    

Other similar striking differences between free and interacting behavior of the entropy are the renormalization of the area term due to mass scales in the theory \cite{area,metli,smolkin}, and notably relative entropy for two states in the limit where the region is a null surface \cite{null}. In all these cases, as in our conjecture here, the apparent paradox of the discontinuity between free and interacting (why the interacting result would not converge to the free one when the charges go to zero?) would in fact be smoothly controlled by a geometric parameter, that here is the separating distance $\epsilon$. However, if one defines the universal term in the entropy as the one resulting from the $\epsilon\rightarrow 0$ limit, then the result would be different for free and interacting models, but for all interesting cases where the gauge field is not completely decoupled the anomaly would be the adequate number to consider.     

\section*{Acknowledgments}
This work was supported by CONICET, CNEA
and Universidad Nacional de Cuyo, Argentina. H.C. acknoledges support from the Simons Foundation through ``It from Qubit'' collaboration.

\appendix
\section{Massless scalar in $d=2$ with Dirichlet boundary condition}
In this appendix we compute analytically the correlation functions on a lattice for a massless scalar in a half line with boundary condition at the origin $\phi(0)=0$. This correlation functions allow us to compute the entropy and mutual information of this mode in the continuum limit with great precision as we can increase the size of the regions without need of dealing with a lattice of finite size. In particular, it is used in the main text to produce the figure (\ref{zeromode}) of the mutual information in this model, and to evaluate the interpolating function $f(\eta)$. 
 
The discrete Hamiltonian is
\be
H=\frac{1}{2}\left(\sum_{i=1}^{\infty} \pi_i^2+(\phi_i-\phi_{i-1})^2\right)\,,
\ee
with $\phi_0=0$. Then, the matrix $K$ is
\be
K_{ij}=2\delta_{ij}-\delta_{j,i+1}-\delta_{j,i-1}\,.
\ee
To evaluate the entropy, we need the two point functions $X=\frac{1}{2\sqrt{K}}$ and $P=\frac{\sqrt{K}}{2}$. 

The normalized eigenstates of $K$ are $\psi_k^l=\sqrt{2/\pi}\sin(kl)$ since we have
\be
\sum_l K_{i l} \sin(k l)=(2-2\cos(k))\sin(k i)\,,
\ee
with $k\in[0,\pi]$. 
From this we obtain
\bea
X_{ij}&=&\frac{1}{\sqrt{2}\pi}\int_0^{\pi}dk \frac{\sin(k i)\sin(kj)}{\sqrt{1-\cos(k)}}\\
&&\hspace{1.7cm}= \frac{1}{4\pi}\left(\psi[1/2-i-j]+\psi[1/2+i+j]-\psi[1/2+i-j]-\psi[1/2-i+j]\right)\,,\nonumber 
\eea
and
\be
P_{ij}=\frac{\sqrt{2}}{\pi}\int_0^{\pi}dk \sin(k i)\sin(kj)\sqrt{1-\cos(k)}=-\frac{4ij}{\pi(2(i^4+j^4)-(i^2+j^2)-4i^2j^2+\frac{1}{8})}\,,
\ee 
where $\psi[x]$ is the digamma function.


\begin{thebibliography}{99}

\bibitem{solodukin} N. Solodukhin, ``Entanglement entropy, conformal invariance and extrinsic geometry,'' Phys. Lett.
B 665, 305 (2008) [arXiv:0802.3117 [hep-th]].


\bibitem{casinihuertamyers}
  H.~Casini, M.~Huerta and R.~C.~Myers,
  ``Towards a derivation of holographic entanglement entropy,''
  JHEP {\bf 1105}, 036 (2011)
  [arXiv:1102.0440 [hep-th]].

\bibitem{casinihuerta}  H.~Casini and M.~Huerta,
  ``Entanglement entropy for the n-sphere,''
  Phys.\ Lett.\ B {\bf 694}, 167 (2010)
  [arXiv:1007.1813 [hep-th]].
  
\bibitem{dowker}J. S.~ Dowker, ``Entanglement entropy for even spheres,'' arXiv:1009.3854 [hep-th].

\bibitem{scalar} R.~ Lohmayer, H.~ Neuberger, A.~ Schwimmer and S.~ Theisen,``Numerical determination of entanglement entropy for a sphere,'' Phys. Lett. B 685, 222 (2010) [arXiv:0911.4283 [hep-lat]]. 

\bibitem{donnelly}
  P.~V.~Buividovich and M.~I.~Polikarpov,
  ``Entanglement entropy in lattice gauge theories,''
  PoS Confinement {\bf 8}, 039 (2008)
  [J.\ Phys.\ A {\bf 42}, 304005 (2009)]
  [arXiv:0811.3824 [hep-lat]].
 W.~Donnelly,
  ``Decomposition of entanglement entropy in lattice gauge theory,''
  Phys.\ Rev.\ D {\bf 85}, 085004 (2012)
  [arXiv:1109.0036 [hep-th]].


 \bibitem{gauge} H.~Casini, M.~Huerta and J.~A.~Rosabal,
  ``Remarks on entanglement entropy for gauge fields,''
  Phys.\ Rev.\ D {\bf 89}, no. 8, 085012 (2014)
  [arXiv:1312.1183 [hep-th]].
  
\bibitem{maxwell} H.~Casini and M.~Huerta,
  ``Entanglement entropy for a Maxwell field: Numerical calculation on a two dimensional lattice,''
  Phys.\ Rev.\ D {\bf 90}, no. 10, 105013 (2014)
  [arXiv:1406.2991 [hep-th]].
  
  \bibitem{lattice}
   R.~M.~Soni and S.~P.~Trivedi,
  ``Aspects of Entanglement Entropy for Gauge Theories,''
  arXiv:1510.07455 [hep-th];
   K.~Van Acoleyen, N.~Bultinck, J.~Haegeman, M.~Marien, V.~B.~Scholz and F.~Verstraete,
  ``The entanglement of distillation for gauge theories,''
  arXiv:1511.04369 [quant-ph];
  S.~Ghosh, R.~M.~Soni and S.~P.~Trivedi,
  ``On The Entanglement Entropy For Gauge Theories,''
  JHEP {\bf 1509}, 069 (2015)
  [arXiv:1501.02593 [hep-th]];
  	L.~Y.~Hung and Y.~Wan,
  ``Revisiting Entanglement Entropy of Lattice Gauge Theories,''
  JHEP {\bf 1504}, 122 (2015)
  [arXiv:1501.04389 [hep-th]];
	S.~Aoki, T.~Iritani, M.~Nozaki, T.~Numasawa, N.~Shiba and H.~Tasaki,
  ``On the definition of entanglement entropy in lattice gauge theories,''
  JHEP {\bf 1506}, 187 (2015)
  [arXiv:1502.04267 [hep-th]];
	  D.~Radicevic,
  ``Entanglement in Weakly Coupled Lattice Gauge Theories,''
  arXiv:1509.08478 [hep-th].
  
  
 \bibitem{Huang} 
  K.~W.~Huang,
  ``Central Charge and Entangled Gauge Fields,''
  Phys.\ Rev.\ D {\bf 92}, no. 2, 025010 (2015)
  [arXiv:1412.2730 [hep-th]].

\bibitem{DonnellyWall} 
  W.~Donnelly and A.~C.~Wall,
  ``Entanglement entropy of electromagnetic edge modes,''
  Phys.\ Rev.\ Lett.\  {\bf 114}, no. 11, 111603 (2015)
  [arXiv:1412.1895 [hep-th]].
\bibitem{theisen} 
  C.~Eling, Y.~Oz and S.~Theisen,
  ``Entanglement and Thermal Entropy of Gauge Fields,''
  JHEP {\bf 1311}, 019 (2013)
  [arXiv:1308.4964 [hep-th]].


\bibitem{bunch} See for example: W.~Donnelly and A.~C.~Wall,
  ``Geometric entropy and edge modes of the electromagnetic field,''
  arXiv:1506.05792 [hep-th];
	 C.~P.~Herzog, K.~W.~Huang and K.~Jensen,
  ``Universal Entanglement and Boundary Geometry in Conformal Field Theory,''
  arXiv:1510.00021 [hep-th];
	 D.~V.~Fursaev,
  ``Entanglement Renyi Entropies in Conformal Field Theories and Holography,''
  JHEP {\bf 1205}, 080 (2012)
  [arXiv:1201.1702 [hep-th]];
	 W.~Donnelly and A.~C.~Wall,
  ``Do gauge fields really contribute negatively to black hole entropy?,''
  Phys.\ Rev.\ D {\bf 86}, 064042 (2012)
 [arXiv:1206.5831 [hep-th]];
	 L.~De Nardo, D.~V.~Fursaev and G.~Miele,
  ``Heat kernel coefficients and spectra of the vector Laplacians on spherical domains with conical singularities,''
  Class.\ Quant.\ Grav.\  {\bf 14}, 1059 (1997)
  [hep-th/9610011];
   D.~N.~Kabat,
  ``Black hole entropy and entropy of entanglement,''
  Nucl.\ Phys.\ B {\bf 453}, 281 (1995)
   [hep-th/9503016].
 

\bibitem{Srednicki}
 M.~Srednicki,
 ``Entropy and area,''
 Phys.\ Rev.\ Lett.\  {\bf 71}, 666 (1993)
 [arXiv:hep-th/9303048].

\bibitem{review} H.~Casini and M.~Huerta,
  ``Entanglement entropy in free quantum field theory,''
  J.\ Phys.\ A {\bf 42}, 504007 (2009)
  [arXiv:0905.2562 [hep-th]].
\bibitem{scalarfield} H.~Casini and M.~Huerta,
  ``Entanglement and alpha entropies for a massive scalar field in two dimensions,''
  J.\ Stat.\ Mech.\  {\bf 0512}, P12012 (2005)
  [cond-mat/0511014].

\bibitem{area} H.~Casini, F.~D.~Mazzitelli and E.~Teste,
  ``Area terms in entanglement entropy,''
  Phys.\ Rev.\ D {\bf 91}, no. 10, 104035 (2015)
  [arXiv:1412.6522 [hep-th]].
   \bibitem{cc}P.~Calabrese and J.~L.~Cardy,
  ``Entanglement entropy and quantum field theory,''
  J.\ Stat.\ Mech.\  {\bf 0406}, P06002 (2004)
  doi:10.1088/1742-5468/2004/06/P06002
  [hep-th/0405152].
 
  
\bibitem{fmutual} 
  H.~Casini, M.~Huerta, R.~C.~Myers and A.~Yale,
  ``Mutual information and the F-theorem,''
  JHEP {\bf 1510}, 003 (2015)
  doi:10.1007/JHEP10(2015)003
  [arXiv:1506.06195 [hep-th]].
  

\bibitem{latorre}  A.~Riera and J.~I.~Latorre,
  ``Area law and vacuum reordering in harmonic networks,''
  Phys.\ Rev.\ A {\bf 74}, 052326 (2006)
  doi:10.1103/PhysRevA.74.052326
  [quant-ph/0605112].
  
\bibitem{Astaneh} 
  A.~F.~Astaneh, A.~Patrushev and S.~N.~Solodukhin,
  ``Entropy vs Gravitational Action: Do Total Derivatives Matter?,''
  arXiv:1411.0926 [hep-th].
  
\bibitem{herait} J.~Lee, A.~Lewkowycz, E.~Perlmutter and B.~R.~Safdi,
  ``Renyi entropy, stationarity, and entanglement of the conformal scalar,''
  arXiv:1407.7816 [hep-th];
C.~P.~Herzog,
  ``Universal Thermal Corrections to Entanglement Entropy for Conformal Field Theories on Spheres,''
  JHEP {\bf 1410}, 28 (2014)
  [arXiv:1407.1358 [hep-th]].
  
\bibitem{metli}Max A. Metlitski, Carlos A. Fuertes, Subir Sachdev, ``Entanglement Entropy in the O(N) model,'' Physical Review {\bf B80}, 115122 (2009) arXiv:0904.4477
\bibitem{smolkin} 
  L.~Y.~Hung, R.~C.~Myers and M.~Smolkin,
  ``Some Calculable Contributions to Holographic Entanglement Entropy,''
  JHEP {\bf 1108}, 039 (2011)
  doi:10.1007/JHEP08(2011)039
  [arXiv:1105.6055 [hep-th]].
\bibitem{null}   R.~Bousso, H.~Casini, Z.~Fisher and J.~Maldacena,
  ``Entropy on a null surface for interacting quantum field theories and the Bousso bound,''
  Phys.\ Rev.\ D {\bf 91}, no. 8, 084030 (2015)
  doi:10.1103/PhysRevD.91.084030
  [arXiv:1406.4545 [hep-th]].
	
  
\end{thebibliography}
\end{document}